\documentclass[journal,twoside]{IEEEtran}

\usepackage{makeidx,epsfig}
\usepackage{setspace,graphicx,multirow}

\usepackage{amsfonts,latexsym,amssymb,amsthm}

\usepackage{graphicx}
\usepackage{epstopdf}

\usepackage{caption3} 
\DeclareCaptionOption{parskip}[]{} 
\usepackage[small]{caption}

\usepackage{subcaption}

\usepackage{array}

\usepackage[cmex10]{amsmath}

\usepackage{url}

\usepackage{multirow}
\usepackage{hhline}

\usepackage{amsmath}
\usepackage{algorithm}
\usepackage[]{algpseudocode}
\usepackage{varwidth}

\makeatletter
\def\BState{\State\hskip-\ALG@thistlm}
\makeatother

\usepackage{algcompatible}

\usepackage{fixltx2e}

\newcommand*\diff{\mathop{}\!\mathrm{d}}

\usepackage{color}

\begin{document}
\bibliographystyle{IEEEtran}

\title{Cyclical Multiple Access in UAV-Aided Communications: A Throughput-Delay Tradeoff}

\author{Jiangbin~Lyu,~\textit{Member,~IEEE},
        Yong~Zeng,~\textit{Member,~IEEE},
        and~Rui~Zhang,~\textit{Senior~Member,~IEEE}%
\thanks{The authors are with the Department of Electrical and Computer Engineering, National University of Singapore (email: jiangbin.lu@u.nus.edu, elezeng@nus.edu.sg, elezhang@nus.edu.sg).}%
\vspace{-3ex}}

\maketitle

\begin{abstract}
This letter studies a wireless system consisting of distributed ground terminals (GTs) communicating with an unmanned aerial vehicle (UAV) that serves as a mobile base station (BS).
The UAV flies cyclically above the GTs at a fixed altitude, which results in a cyclical pattern of the strength of the UAV-GT channels.
To exploit such periodic channel variations, we propose a new \textit{cyclical multiple access (CMA)} scheme to schedule the communications between the UAV and GTs in a cyclical time-division manner based on the flying UAV's position.
The time allocations to different GTs are optimized to maximize their minimum throughput. It is revealed that there is a fundamental tradeoff between throughput and access delay in the proposed CMA.
Simulation results show significant throughput gains over the case of a static UAV BS in delay-tolerant applications.
\end{abstract}

\begin{IEEEkeywords}
Unmanned aerial vehicles, mobile base station, periodic channel, cyclical multiple access, throughput-delay tradeoff
\end{IEEEkeywords}

\vspace{-3ex}
\section{Introduction}

%

With their high mobility and reducing cost, unmanned aerial vehicles (UAVs) have found applications in wireless communication systems \cite{ZengUAVmag}, either to support the existing cellular networks in high-demand and overloaded situations \cite{UAVoverload}, or to provide wireless connectivity in areas without infrastructure coverage such as battle fields or disaster scenes \cite{UAVpublicSafety}.
Compared to terrestrial communications, UAV-aided wireless systems are in general faster to deploy, more flexibly reconfigured, and likely to have better communication channels due to line-of-sight links.
For example,
the UAV can be deployed as a quasi-stationary aerial base station (BS) for the ground terminals (GTs) \cite{LAPlosProbability,DroneSmallCell}. Thanks to high mobility, UAVs can also be deployed as mobile relays to provide wireless connectivity between distant GTs whose direct links are severely blocked  \cite{UAVrelay,ZengMobileRelay}.
The deployment of UAVs is also considered in \cite{HanZhuUAV} to improve connectivity of wireless ad-hoc networks.

In this letter, we consider a wireless system with the UAV deployed as a mobile BS to provide wireless connectivity for a group of distributed GTs.
Compared to static UAV BSs in \cite{LAPlosProbability,DroneSmallCell}, the UAV controlled mobility is exploited to improve channel quality and thus system throughput.
We assume that the UAV flies cyclically above the GTs at a fixed altitude, which results in cyclically varying patterns of the strength of the UAV-GT channels. We thus propose a new \textit{cyclical multiple access (CMA)} scheme where GTs are scheduled to communicate with the UAV
in a cyclical time-division manner to exploit the good channel when the UAV flies closer to each of them.
We propose an algorithm to allocate the time to different GTs based on the flying UAV's position to maximize their minimum throughput. We also analyze the access delay patterns of the GTs due to CMA and reveal a fundamental trade-off between throughput and delay.
Numerical results show that the max-min throughput is significantly improved by the proposed CMA with a mobile UAV BS over the case of a static BS, at the expense of increased access delay.
Thus, the proposed UAV-assisted mobile BS with CMA is most suitable in delay-tolerant applications \cite{DTNsurvey} such as periodic sensing, large data transfer, etc.
The mobile UAV BS case is also considered in \cite{UAVunderlayD2D} to optimize the coverage performance.





\section{System Model}\label{SectionModel}
\begin{figure}
\centering
\includegraphics[width=0.85\linewidth,  trim=0 0 0 0,clip]{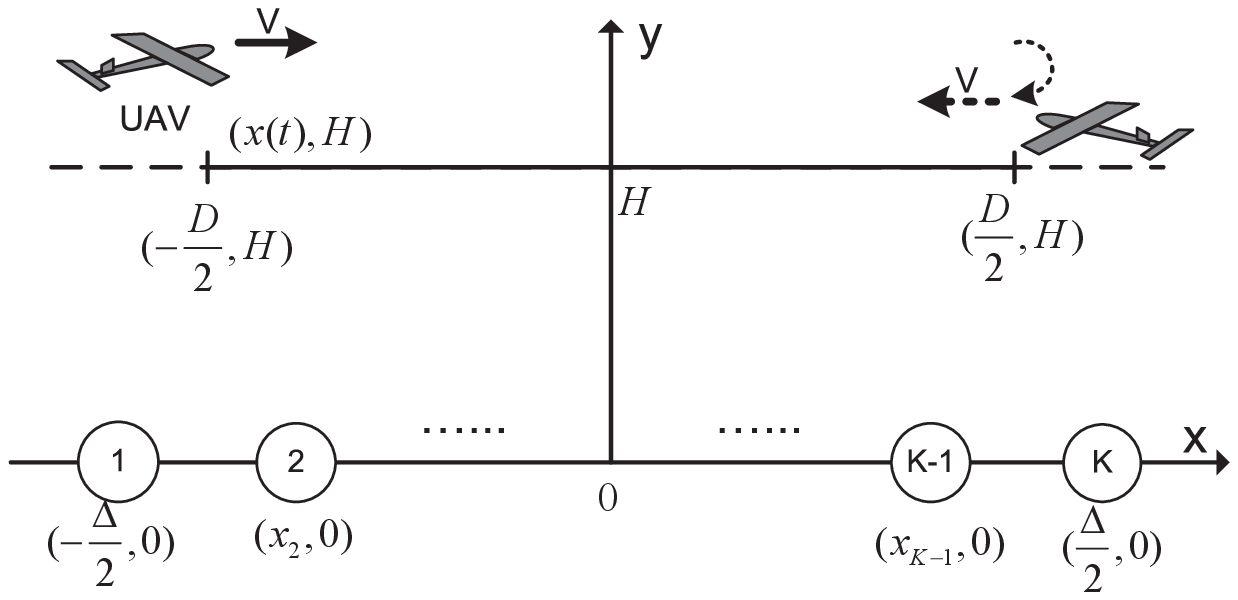} 
\caption{A wireless communication system with a UAV-aided mobile BS moving along a straight line.\vspace{-3ex}} \label{UAV2DtInterval}
\end{figure}

We consider a wireless communication system where a UAV is employed as a mobile BS to communicate with $K$ GTs. We assume that the GTs are equally spaced on the ground along a straight line with the length of $\Delta$ meters, as shown in Fig.~\ref{UAV2DtInterval}. Thus, the location $x_k$ of the $k$th GT can be expressed as $x_k=-\Delta/2+(k-1)\Delta/(K-1)$, $k=1,\cdots, K$. Note that the assumption of one-dimensional (1D) uniform GT locations is for the purpose of exposition to draw essential insights to the performance and design of the new UAV-assisted communication system. The results can be extended to arbitrary GT locations in general 2D or 3D setups, which are left for our future work. In practice, each GT in the considered 1D model could correspond to a cluster head that serves as a gateway for a cluster of nearby nodes communicating with the UAV. When the nodes are densely populated and the number of clusters is large, the cluster heads can be approximated to be equally spaced.

We assume that the UAV flies horizontally above the GTs following a cyclical trajectory with period $T$, i.e., the UAV position repeats every $T$ seconds. 
For simplicity, we assume that the UAV flies at a fixed altitude $H$, which could correspond to the minimum value required for safety considerations (e.g., terrain or building avoidance). The extension to variable $H$ will be left as future work.
For the symmetric 1D GT distribution considered in this letter, it is intuitive that the UAV's trajectory $x(t), 0\leq t\leq T$, should be symmetric over the origin. Thus, assuming the UAV flies with a constant speed $V$, we have $x(t)\in[-D/2,D/2]$, $\forall 0\leq t\leq T$, with the trajectory length $D=VT/2$, as shown in Fig. \ref{UAV2DtInterval}.

We consider the downlink communications where the UAV sends independent messages to the GTs over a given frequency band, whereas the obtained results can be similarly applied to uplink communications. We assume the communication channels are dominated by line-of-sight (LoS) links, and the Doppler effect due to the UAV's mobility is assumed to be perfectly compensated.
Though simplified, the LoS model 
offers a good approximation for the UAV-GT channels in practice, which suffices for us to characterize the fundamental performance tradeoff.
Therefore, at time instant $t$, the channel power gain from the UAV to each GT $k$ follows the free-space path loss model given by
\begin{equation}\label{PathLoss}
h_k(t)=\beta_0 d_k^{-2}(t)=\frac{\beta_0}{(x(t) - x_k)^2+H^2}, 0\leq t\leq T,
\end{equation}
where $\beta_0$ denotes the channel power gain at the reference distance $d_0=1$ meter (m), whose value depends on the carrier frequency, antenna gain, etc., and $d_k(t)=\sqrt{(x(t) - x_k)^2+H^2}$ is the link distance between the UAV and the $k$-th GT at time $t$.
Note that from (\ref{PathLoss}), it follows that for each GT $k$ with fixed location $x_k$, its channel with the moving UAV with position $x(t)$ varies in a periodic manner over $T$, which is perfectly known at the UAV.
By assuming a constant transmission power $P$ by the UAV, the instantaneous channel capacity from the UAV to the $k$-th GT in bits/second/Hz (bps/Hz) can be expressed as
\begin{equation}\label{rkt}
r_k(t)=\log_2\bigg(1+\frac{P h_k(t)}{\sigma^2}\bigg), 0\leq t\leq T,
\end{equation}
where $\sigma^2$ denotes the noise power at each GT receiver.
In the following, we use the unit of bps/Hz to measure the throughput per unit bandwidth, also known as the spectrum efficiency.

\section{Cyclical Multiple Access}\label{CyclicalMultipleAccess}

To exploit periodic channel variations of different GTs, we propose a new multiple access scheme called CMA where GTs communicate with the UAV in a cyclical time-division manner, for which the details are given in this section.

\subsection{Cyclical TDMA}
\begin{figure}
\centering
\includegraphics[width=1\linewidth,  trim=80 0 50 0,clip]{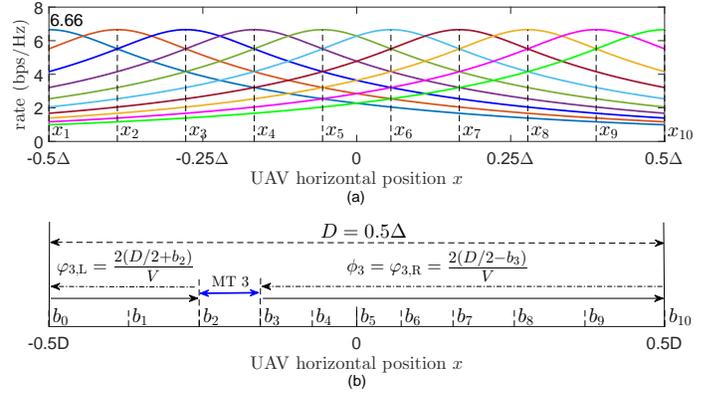} 
\caption{Cyclical multiple access with 10 GTs: (a) UAV-GT rate distribution; (b) transmission time allocation for different GTs.\vspace{-3ex}} \label{K10plot}
\end{figure}

From \eqref{PathLoss} and \eqref{rkt}, the maximum achievable rate $r_k$ of GT $k$ as a function of the UAV horizontal position $x$ is given by
\begin{equation}\label{rkx}
r_k(x)=\log_2\bigg(1+\frac{P\gamma_0}{(x - x_k)^2+H^2}\bigg), k=1,\cdots,K,
\end{equation}
where $\gamma_0\triangleq \beta_0/\sigma^2$ represents the reference signal-to-noise ratio (SNR).
From \eqref{rkx}, it follows that for each GT $k$, the rate $r_k$ is symmetric and unimodal, which achieves its maximum at $x=x_k$. Moreover, the rate expressions of different GTs are identical except that they have different shifts along the x-coordinate according to different GT locations $x_k$'s .
As an illustration, Fig.~\ref{K10plot}(a) plots the instantaneous rate of each GT versus the UAV position $x$, with $K=10$, $P=10$dBm, $\gamma_0=80$dB, $H=100$m, and $\Delta=1000$m.

Since each GT has its highest rate $r_k$ when $x=x_k$, it is intuitive to allocate the segment of the UAV trajectory near $x=x_k$ to GT $k$ for communication, so as to maximize the throughput to all GTs.
Motivated by this, we propose a simple time-division based CMA, termed \textit{cyclical TDMA}, to schedule the communications from the UAV to different GTs based on the UAV position. Specifically,
the one-way UAV trajectory of length $D$ is divided into $K$ contiguous horizontal segments: $[-D/2,b_1], [b_1,b_2], \cdots, [b_{K-1},D/2]$, which are allocated to the $K$ GTs for orthogonal transmissions, respectively.
For convenience, let $b_0=-D/2$ and $b_K=D/2$.
Then the UAV trajectory segment $[b_{k-1},b_{k}]$ and its corresponding time interval $[t_{k-1},t_{k}]$ is allocated to GT $k$.
With such a periodic TDMA scheme, the average throughput of GT $k$ in bps/Hz is obtained as
\begin{equation}\label{Throughput}
\theta_k=\frac{2}{T}\int_{t_{k-1}}^{t_k} r_k(t)\diff t=\frac{2}{T}\int_{b_{k-1}}^{b_k} r_k(x)\diff x \frac{\diff t}{\diff x}=\frac{1}{D}\int_{b_{k-1}}^{b_k} r_k(x)\diff x,
\end{equation}
where we have used the identity $\frac{\diff t}{\diff x}=1/V$ and $D=VT/2$. 
The integral of the rate function $r_k(x)$ can be obtained in closed-form as
\begin{multline}\label{rkIntegral}
R_k (x)=\int r_k(x)\diff x=(x-x_k)\log_2\bigg(1+\frac{P\gamma_0}{(x - x_k)^2+H^2}\bigg)\\
+2\bigg(H\tan^{-1}\frac{x_k-x}{H} -\sqrt{H^2+P\gamma_0}\tan^{-1}\frac{x_k-x}{\sqrt{H^2+P\gamma_0}}\bigg)/\ln 2.
\end{multline}
Therefore, the throughput of GT $k$ is given by
\begin{equation}\label{Throughputb}
\theta_k=\frac{1}{D}R_k(x)|_{b_{k-1}}^{b_k}=\frac{1}{D}\bigg(R_k(b_k)-R_k(b_{k-1})\bigg).
\end{equation}

\subsection{Max-Min Throughput}\label{SectionFixedTraj}
For a fixed UAV trajectory length $D$, we study the problem of maximizing the minimum throughput among all GTs, denoted by $\tau$, by optimizing the delimiting variables $b_1,b_2,\cdots,b_{K-1}$ for their time allocations. 
The max-min throughput is used to maximize the overall throughput while ensuring fairness among all GTs.
The problem can be formulated as
\begin{align}
\mathrm{(P1)}:
\begin{cases}
\underset{\{b_k\}_{k=1}^{K-1}}{\max} & \ \tau   \notag \\
\text{s.t.}  & \ \theta_k\geq \tau, k=1,\cdots, K,\\
 & \ -D/2\leq b_1\leq b_2\leq\cdots\leq b_{K-1}\leq D/2,
\end{cases}
\end{align}%
where $\theta_k$ is a non-concave function of $b_{k-1}$ and $b_k$ from  \eqref{rkIntegral} and \eqref{Throughputb}.
Therefore, (P1) is non-convex and thus cannot be directly solved via standard convex optimization techniques.

Fortunately, (P1) can be efficiently solved by exploiting the following two properties.
Firstly, it can be proved that all GTs have equal throughput when the max-min throughput $\tilde\tau$ is achieved.
Secondly,
note that if all other delimiting variables are fixed, the optimization variable $b_k$ only affects the throughputs of GT $k$ and GT $k+1$.
The first property can be proved by contradiction.
Assume that the max-min throughput $\tilde\tau$ is achieved at a single GT $k$ while its neighboring GT $k+1$ (or $k-1$ if $k=K$) has a higher throughput,
i.e., 
\begin{equation}\label{ThroughputbCompare}
\tilde\tau=\tilde\theta_k=\frac{1}{D}R_k(x)|_{\tilde b_{k-1}}^{\tilde b_{k}}<\frac{1}{D}R_{k+1}(x)|_{\tilde b_{k}}^{\tilde b_{k+1}}=\tilde\theta_{k+1}.
\end{equation}
As $b_k$ increases from $\tilde b_{k-1}$ to $\tilde b_{k+1}$, $\theta_k$ monotonically increases from $0$ while $\theta_{k+1}$ monotonically decreases to $0$.
Therefore, there exists a unique value of $b_k=\hat b_k>\tilde b_{k}$ which makes 
\begin{equation}\label{ThroughputbCompare1}
\hspace*{-0.3cm}\hat\theta_{k+1}=\frac{1}{D}R_{k+1}(x)|_{\hat b_{k}}^{\tilde b_{k+1}}=\hat\theta_k=\frac{1}{D}R_k(x)|_{\tilde b_{k-1}}^{\hat b_{k}}>\tilde\theta_k=\tilde\tau.
\end{equation}
This contradicts with the assumption that $\tilde\tau$ is the max-min throughput, since it can be increased to $\hat\theta_k$. The cases in which the max-min throughput $\tilde\tau$ is achieved at multiple GTs can be proved in a similar way.

The above proof also suggests a method to obtain the max-min throughput $\tilde\tau$.
Based on the above two properties, we can fix $b_1,\cdots,b_{k-1},b_{k+1},\cdots,b_{K-1}$ and optimize $b_k$ to ensure that $\theta_k=\theta_{k+1}$, i.e., $b_k$ can be updated by solving for $b$ from the following equation:
\begin{equation}\label{ThroughputEqual}
R_k(b)-R_k(b_{k-1})=R_{k+1}(b_{k+1})-R_{k+1}(b), b_{k-1}\leq b\leq b_{k+1}.
\end{equation}
We therefore propose Algorithm \ref{Alg1} to iteratively tune the delimiting variables to achieve the max-min throughput for (P1). In each iteration, we pick the delimiting variable $b_{k_0}$ that corresponds to the largest throughput gap $\zeta_{max}$ between two neighboring GTs $k_0$ and $k_0+1$. Then we tune $b_{k_0}$ to eliminate this gap by solving equation (\ref{ThroughputEqual}). The iterations repeat until all the throughput gaps between two neighboring GTs are smaller than a certain threshold $\epsilon$.

\begin{algorithm}[H]\caption{Max-Min Throughput with Cyclical TDMA}\label{Alg1}
\begin{small}
\textbf{Input:} GT locations $x_k, 1\leq k\leq K$; UAV altitude $H$, trajectory length $D$; transmit power $P$, reference SNR $\gamma_0$.\\
\textbf{Output:} Trajectory delimiting variables $b_1,\cdots, b_{K-1}$; max-min throughput $\tilde\tau$.\\
\textbf{Initialization:} $b_0=-D/2$, $b_K=D/2$, $b_k=b_0+k(D/K)$, $1\leq k\leq K-1$; $\epsilon=10^{-5}$.
\begin{algorithmic}[1]
\REPEAT
\STATE Calculate throughput $\theta_k,\forall k$ from (\ref{Throughputb}).
\STATE Find the largest throughput gap $\zeta_{max}=\max\limits_{1\leq k\leq K-1}|\theta_{k+1}-\theta_k|$ and corresponding index $k_0$.
\STATE Solve for $b$ in equation (\ref{ThroughputEqual}) with $k=k_0$. Update $b_{k_0}=b$.
\UNTIL{$\zeta_{max}<\epsilon$}
\STATE The max-min throughput $\tilde\tau=\theta_1$.
\end{algorithmic}
\end{small}
\end{algorithm}


Algorithm \ref{Alg1} achieves the max-min throughput $\tilde\tau$ for any fixed trajectory length $D$, which can be proved by contradiction.
First of all, Algorithm \ref{Alg1} is guaranteed to converge since the largest throughput gap $\zeta_{max}$ is bounded below by 0 and is decreasing in each iteration.
After convergence,
Algorithm \ref{Alg1} returns the delimiting variables $\tilde b_1,\cdots, \tilde b_{K-1}$ which yields
$\zeta_{max}\rightarrow 0$ and thus all GTs have equal throughputs given by
\begin{equation}\label{ThroughputbContradict}
\frac{1}{D}R_1(x)|_{-\frac{D}{2}}^{\tilde b_{1}}=\frac{1}{D}R_2(x)|_{\tilde b_{1}}^{\tilde b_{2}}=\cdots=\frac{1}{D}R_K(x)|_{\tilde b_{K-1}}^{\frac{D}{2}}=\tilde\tau,
\end{equation}
where $\tilde\tau$ denotes the achievable max-min throughput. Now assume there exists a higher max-min throughput $\hat\tau>\tilde\tau$, with the delimiting variables $\hat b_1,\cdots, \hat b_{K-1}$.
Then we have
\begin{equation}\label{ThroughputbCompare}
\frac{1}{D}R_1(x)|_{-\frac{D}{2}}^{\hat b_{1}}\geq\hat\tau>\tilde\tau=\frac{1}{D}R_1(x)|_{-\frac{D}{2}}^{\tilde b_{1}}.
\end{equation}
Since the integral function $R_1(x)$ is monotonically increasing, from (\ref{ThroughputbCompare}) we have $\hat b_{1}>\tilde b_{1}$.
By induction, we can similarly conclude that $\hat b_{2}>\tilde b_{2}, \cdots, \hat b_{K-1}>\tilde b_{K-1}$. However, this contradicts with the assumption that $\hat\tau>\tilde\tau$, which also results in $\hat b_{K-1}<\tilde b_{K-1}$ from the following inequality:
\begin{equation}\label{ThroughputbCompare2}
\frac{1}{D}R_K(x)|_{\hat b_{K-1}}^{\frac{D}{2}}\geq\hat\tau>\tilde\tau=\frac{1}{D}R_K(x)|_{\tilde b_{K-1}}^{\frac{D}{2}}.
\end{equation}
Therefore, $\tilde\tau$ achieved by Algorithm \ref{Alg1} is indeed the max-min throughput. 
The proof is thus completed.$\blacksquare$

For the example in Fig. \ref{K10plot}(a) with $K=10$, assume that the one-way trajectory length is set to $D=\Delta/2$.
Denote the portion of the UAV trajectory allocated to GT $k$ as $\delta_k$, i.e.,
\begin{equation}\label{portion}
\delta_k=\frac{b_{k}-b_{k-1}}{D}.
\end{equation}
Algorithm \ref{Alg1} is applied to obtain the optimal delimiting variables $b_1,\cdots, b_{9}$ shown in Fig. \ref{K10plot}(b), which achieve the max-min throughput $\tilde{\tau}=0.4663$ bps/Hz.
Note that the allocated trajectory portions are non-uniform for different GTs.
In general, the middle GTs require a smaller portion of the trajectory to achieve the same throughput since they enjoy better channels than the GTs on the two sides.

For benchmark comparison with a static UAV BS, since all GTs are symmetric around the origin, the optimal fixed UAV position for max-min throughput is $(0,H)$ due to symmetry.
Since the UAV-GT channels are fixed, by optimizing the time allocation, the max-min throughput $\tilde\tau$ in the static UAV case can be obtained as
\begin{equation}\label{Tauc}
\tilde\tau=\frac{1}{\sum_{k=1}^K 1/ r_k},
\end{equation}
which has a value of 0.3488 bps/Hz for the example in Fig. \ref{K10plot}(a).
The static UAV BS scenario can be treated as an extreme case of the proposed cyclical TDMA with zero UAV trajectory length, i.e., $D=0$.
It is found that the max-min throughput achieved by the mobile UAV BS with $D=\Delta/2$ is 33.7\% higher than that of the static UAV BS case.
This gain is owing to the UAV mobility, which can be further improved by optimizing the trajectory length $D$ (as will be shown later in Section \ref{SectionSimu}).
However, for a fixed UAV speed $V$, the trajectory period $T=2D/V$ and hence the access delay of the GT communications increases with $D$.
Therefore, there exists a general tradeoff between maximizing the throughput and minimizing the access delay, in selecting the value of $D$ in the proposed mobile UAV BS with CMA.

\subsection{Access Delay}\label{SectionDelay}
Within each period $T$ of the UAV cyclical trajectory from $x=-D/2$ to $D/2$ and then back to $-D/2$, each GT $k$ can communicate with the UAV in two non-consecutive time windows corresponding to the UAV position $x\in[b_{k-1},b_{k}]$. This results in two non-consecutive ``mute" time windows in which GT $k$ cannot communicate with the UAV whose lengths are denoted as $\varphi_{k,\rm L}$ and $\varphi_{k,\rm R}$, corresponding to the UAV trajectory on the left and right sides, respectively.
For the example with 10 GTs, the UAV trajectory segment $x\in[b_2,b_3]$ is allocated to GT 3, with $\varphi_{3,\rm L}$ and $\varphi_{3,\rm R}$ also shown in Fig. \ref{K10plot}(b).

Define the access delay $\phi_k$ as the longest contiguous mute time of GT $k$ in a UAV flying period $T$, i.e.,
\begin{equation}\label{Delay}
\phi_k=\max\{\varphi_{k,\rm L},\varphi_{k,\rm R}\}=\max\bigg\{\frac{2(D/2+b_{k-1})}{V}  ,  \frac{2(D/2-b_{k})}{V}\bigg\}.
\end{equation}
According to this definition, the GTs have different access delays depending on their relative locations along the UAV trajectory.
In general, the middle GTs have smaller access delays since $\varphi_{k,\rm L}$ and $\varphi_{k,\rm R}$ are roughly the same, while the GTs on the two sides have larger access delays since $\varphi_{k,\rm L}$ and $\varphi_{k,\rm R}$ are unbalanced.
Such access delay patterns need to be considered when designing upper-layer protocols or applications with different delay requirements.

Note that the access delay depends on the flying speed $V$, UAV trajectory length $D$, as well as the delimiting variables $b_k, 1\leq k\leq K-1$, which need to be obtained using Algorithm \ref{Alg1}.
There could be various ways to characterize the overall access delay of all GTs. In this letter, we adopt the
root-mean-square (RMS) access delay $\phi_{rms}$ defined as
\begin{equation}\label{rmsDelay}
\phi_{rms}=\sqrt{\frac{1}{K}(\phi_1^2+\phi_2^2+\cdots+\phi_K^2)},
\end{equation}
which accounts for both the average value and variations of the delay.
In general, $\phi_{rms}$ monotonically increases with $D$, since the round-trip period $T=2D/V$ is proportional to $D$ with fixed $V$. 
In this letter, we assume the system requires an RMS access delay no larger than $\Phi$, i.e., $\phi_{rms}\leq \Phi$.

Note that the access delay considered in this letter is different from the conventional communication delay.
Access delay is caused by the cyclical TDMA in which the GTs take turns to access the channel when the UAV flies above them.
In the limiting case with a static UAV BS (or equivalently, $D=0$ in the mobile UAV BS case), the access delay can be arbitrarily small since the time frame can be divided into arbitrarily small mini-slots, each assigned to one GT for communication. However, this is not the case in cyclical TDMA with non-zero trajectory length $D$, since
given a finite value of UAV speed $V>0$, each GT $k$ may need to wait a maximum time equal to its access delay $\phi_k$ given in (\ref{Delay}) for any two adjacent communications with the UAV (which in practice can be significantly larger than the conventional delay due to packet transmission).

\section{Numerical Results}\label{SectionSimu}

\begin{figure}
\centering
\includegraphics[width=1\linewidth,  trim=30 0 70 0,clip]{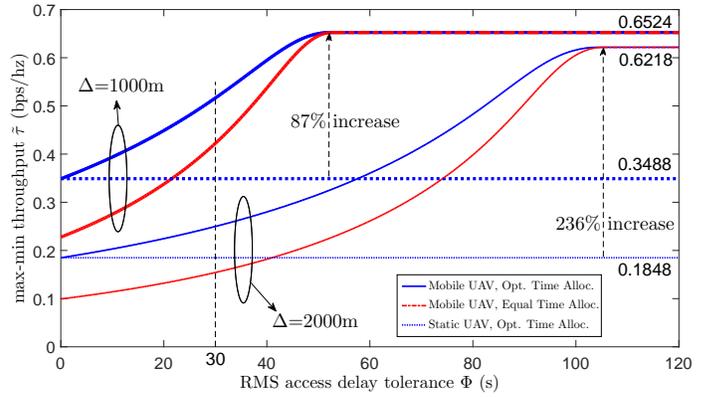} 
\caption{Max-min throughput $\tilde\tau$ versus RMS access delay tolerance $\Phi$} \label{K10Tradeoff}
\vspace{-1em}
\end{figure}

In this section, 
we normalize the trajectory length $D$ to the GT location range $\Delta$ for convenience, denoted as $\bar D=D/\Delta$.
The following parameters are used:
$P=10$ dBm, $\gamma_0=80$ dB, $H=100$ m, and $V=30$ m/s.
For the same example in Fig. \ref{K10plot}(a) with  $K=10$ GTs, 
we compare the max-min throughput $\tilde\tau$ under different RMS access delay tolerance values $\Phi$ to show the fundamental throughput-delay trade-off. For each pair of their values shown in Fig. \ref{K10Tradeoff}, the corresponding optimal $\bar D$ is obtained by searching $\bar D$ subject to that the resulted RMS access delay $\phi_{rms}$ is no larger than the given tolerance value $\Phi$, as follows. First, based on Algorithm \ref{Alg1}, we obtain the max-min throughput $\tilde\tau$ among all GTs for a fixed value of $\bar D$. Then, we find the optimal $\bar D$ such that the resulted RMS delay $\phi_{rms}$ satisfies the given $\Phi$ (i.e., $\phi_{rms}\leq \Phi$) to maximize $\tilde\tau$. This can be done by a simple one-dimensional search as the RMS access delay $\phi_{rms}$ in general monotonically increases with $\bar D$. 
Besides the optimal time/segment allocation proposed in Algorithm \ref{Alg1}, we also consider a simple equal time allocation scheme for comparison, which corresponds to $\delta_k=1/K, \forall k$, with $\delta_k$ similarly defined as in (\ref{portion}).


First, it is observed that for the case of $\Delta=1000$m, the two schemes with optimal and equal time allocations achieve their peak values $\tilde\tau^*=0.6524$ bps/Hz and 0.6523 bps/Hz at $\bar D^*=1.10$ and 1.11, respectively, and the corresponding RMS delays are $\phi_{rms}=52.37$s and 52.85s, respectively.
Therefore, if sufficiently large access delay can be tolerated, e.g., $\Phi>60$s, the simple equal time allocation with the optimal trajectory length $\bar D^*=1.11$ achieves the near-optimal performance.
On the other hand, for relatively stringent access delay requirement, e.g., $\Phi<30$s,
the cyclical TDMA with the optimal time allocation by Algorithm~\ref{Alg1} significantly outperforms that with the equal time allocation.

Next, we investigate the throughput gain of the mobile UAV BS over its static UAV BS special case ($\bar D=0$), both with optimal time allocations.
We have shown in Section \ref{SectionFixedTraj} that for fixed trajectory length $\bar D=0.5$, a throughput gain of 33.7\% is achievable by the mobile UAV BS, whereas the gain further increases to 87\% with optimized $\bar D^*=1.10$ as shown in Fig. \ref{K10Tradeoff}, at the cost of increased access delay.
This gain is further increased in the case of a larger GT location range $\Delta$, i.e., up to 236\% when $\Delta=2000$m as shown in Fig. \ref{K10Tradeoff}.
The rationale is that when $\Delta$ increases,
the GTs on the two sides experience even poorer channels in the static UAV BS case, which degrades the max-min throughput significantly.
On the contrary, cyclical TDMA still maintains a high max-min throughput, since all GTs can enjoy good channels in their allocated UAV trajectory segments, if the trajectory length $\bar D$ is optimized with respect to the location range $\Delta$ accordingly.


%
%
%
%


\section{Conclusions}\label{Conclusion}
This letter proposed a new cyclical multiple access scheme in UAV-aided wireless communications,
and characterized the max-min throughput by optimally allocating the transmission time to GTs based on the UAV position.
Simulation results showed significant throughput gains with the proposed design over the static UAV BS case in delay-tolerant scenarios.
Possible extensions of this letter for future work may include the general GT locations in 2D or 3D setups,
variable speed/altitude control of the UAV,
deployment and cooperation of multiple UAVs, etc.


\bibliography{IEEEabrv,BibDIRP}

\begin{thebibliography}{10}
\providecommand{\url}[1]{#1}
\csname url@samestyle\endcsname
\providecommand{\newblock}{\relax}
\providecommand{\bibinfo}[2]{#2}
\providecommand{\BIBentrySTDinterwordspacing}{\spaceskip=0pt\relax}
\providecommand{\BIBentryALTinterwordstretchfactor}{4}
\providecommand{\BIBentryALTinterwordspacing}{\spaceskip=\fontdimen2\font plus
\BIBentryALTinterwordstretchfactor\fontdimen3\font minus
  \fontdimen4\font\relax}
\providecommand{\BIBforeignlanguage}[2]{{%
\expandafter\ifx\csname l@#1\endcsname\relax
\typeout{** WARNING: IEEEtran.bst: No hyphenation pattern has been}%
\typeout{** loaded for the language `#1'. Using the pattern for}%
\typeout{** the default language instead.}%
\else
\language=\csname l@#1\endcsname
\fi
#2}}
\providecommand{\BIBdecl}{\relax}
\BIBdecl

\bibitem{ZengUAVmag}
Y.~Zeng, R.~Zhang, and T.~J. Lim, ``Wireless communications with unmanned
  aerial vehicles: opportunities and challenges,'' \emph{IEEE Commun. Mag.},
  vol.~54, no.~5, pp. 36--42, May 2016.

\bibitem{UAVoverload}
S.~Rohde and C.~Wietfeld, ``Interference aware positioning of aerial relays for
  cell overload and outage compensation,'' in \emph{Proc. IEEE Vehicular
  Technology Conference (VTC)}, Sept. 2012, pp. 1--5.

\bibitem{UAVpublicSafety}
A.~Merwaday and I.~Guvenc, ``{UAV} assisted heterogeneous networks for public
  safety communications,'' in \emph{Proc. IEEE Wireless Commun. Netw. Conf.},
  Mar. 2015, pp. 329--334.

\bibitem{LAPlosProbability}
A.~Al-Hourani, S.~Kandeepan, and S.~Lardner, ``Optimal {LAP} altitude for
  maximum coverage,'' \emph{IEEE Wireless Commun. Lett.}, vol.~3, no.~6, pp.
  569--572, Dec. 2014.

\bibitem{DroneSmallCell}
M.~Mozaffari, W.~Saad, M.~Bennis, and M.~Debbah, ``Drone small cells in the
  clouds: design, deployment and performance analysis,'' in \emph{Proc. IEEE
  GLOBECOM}, Dec. 2015, pp. 1--6.

\bibitem{UAVrelay}
P.~Zhan, K.~Yu, and A.~L. Swindlehurst, ``Wireless relay communications with
  unmanned aerial vehicles: performance and optimization,'' \emph{IEEE Trans.
  Aerosp. Electron. Syst.}, vol.~47, no.~3, pp. 2068--2085, July 2011.

\bibitem{ZengMobileRelay}
Y.~Zeng, R.~Zhang, and T.~J. Lim, ``Throughput maximization for mobile relaying
  systems,'' \emph{submitted to IEEE Trans. Commun., 2016}, (available online
  at arxiv/1604.02517).

\bibitem{HanZhuUAV}
Z.~Han, A.~L. Swindlehurst, and K.~J.~R. Liu, ``Optimization of manet
  connectivity via smart deployment/movement of unmanned air vehicles,''
  \emph{IEEE Trans. Veh. Technol.}, vol.~58, no.~7, pp. 3533--3546, Sept. 2009.

\bibitem{DTNsurvey}
M.~J. Khabbaz, C.~M. Assi, and W.~F. Fawaz, ``Disruption-tolerant networking: a
  comprehensive survey on recent developments and persisting challenges,''
  \emph{IEEE Commun. Surv. Tut.}, vol.~14, no.~2, pp. 607--640, 2012.

\bibitem{UAVunderlayD2D}
M.~Mozaffari, W.~Saad, M.~Bennis, and M.~Debbah, ``Unmanned aerial vehicle with
  underlaid device-to-device communications: Performance and tradeoffs,''
  \emph{IEEE Trans. on Wireless Commun.}, vol.~15, no.~6, pp. 3949--3963, June
  2016.

\end{thebibliography}

\newpage

\end{document}